\documentclass[final]{aipproc}
\layoutstyle{6x9}
\usepackage{amsmath,amssymb} \usepackage{epsfig}
\usepackage{graphics}\input{dvabbr.sty}
\begin{document}

 \title{\bf QCD Effective Couplings in Minkowskian and
 Euclidean Domains}

 \classification{ 11.10.Hi, 12.38.Cy, 12.38.Lg}
\keywords{Non-perturbative QCD, non-power expansion}

\author{D.V. Shirkov \ \ \today}
 {address={Bogoliubov Lab. of Theor. Physics, JINR,
 Dubna, 141980, Russia}}

\begin{abstract}
 We argue for essential upgrading of the defining equations
 (9.5) and (9.6) in Section 9.2. "The QCD coupling ... " of PDG
 review and their use for data analysis in the light of recent
 development of the QCD theory. Our claim is twofold. First,
 instead of universal expression (9.5) for \albars, one should
 use various ghost-free couplings $\alpha_E(Q^2)\,,\,\,
 \alpha_M(s)\,\dots\,$ \ specific for a given physical
 representation. Second, instead of power expansion (9.6) for
 observable, we recommend to use nonpower functional ones over
 particular functional sets $\left\{{\cal A}_k(Q^2)\right\}\,,$
 $\left\{{\mathfrak A}_k(s)\right\}\,\dots\,$ related by
 suitable integral transformations. We remind that use of this
 modified prescription results in a better correspondence of
 reanalyzed low energy data with the high energy ones.
\end{abstract}

\maketitle

\section{\bf 1. PREAMBLE}
The main message consists of two statements:\par
 {\bf A:} Instead of common effective QCD coupling $\albars\,,$
 (with its ghost defect) as, e.g., it is implicitly mentioned
 by eq.(9.5) of PDG review
 \cite{pdg04}, one should use (at least) two {\sf different
 ghost-free forms for QCD effective coupling} $\alpha_E(Q^2)$
 in the Euclidean and $\alpha_M(s)\,$ in the Minkowskian
 (and, possibly, some others) pictures;\par
 {\bf B:} The RG-invariant perturbative expansions for
 observables, like eq.(9.6) in PDG,
 $$  O(\xi)\,=\,o_1\,\albars(\xi)\,+\,o_2\,\albars^2(\xi)
 \,+   \,o_3\:\albars^3(\xi)\,+\,...\, ,  $$
 -- in~powers {\it of the same \albars} in different pictures,
 Euclidean ($\xi=Q^2\,)$ or Minkowskian $(\xi=s\,$) are neither
 based theoretically, nor adequate practically to {low-energy}
 QCD. Instead, one should use {\sf diverse nonpower functional
 expansions} 
 $$  d(Q^2)=\sum_{i\geq 1}\,d_i\acal_i(Q^2)\,,\quad
 r(s)\,=\sum_{i\geq 1}\,d_i\,\agoth_i(s)\,, $$
 (each particular one for a given representation) over
 nonpower sets of ghost-free functions like
 $\left\{\acal_k{(Q^2)}\right\}\,$ in Euclidean and
 $\left\{\agoth_k(s)\right\}\,$ in Minkowskian, mutually
 related by suitable integral transformations. \par

   Below we demonstrate that a reasonable revising of the above
 mentioned PDG Eqs. essentially modifies the results of the
 analysis of some low energy data like GLM and Bjorken sum-rules,
 $\tau$-lepton and Ypsilon decays and $e+e-\,$inclusive
 cross-sections (Sections 9.3., 9.4. and 9.6 in PDG).\par
   As a result, new overall fit for Euclidean data in terms of
 $\alpha_E(Q^2)\,$ and Minkowskian data in $\alpha_M(s)\,$
 results in (see our recent review \cite{epj01})
 $\albars(M_Z^2)= 0.123\,$ with an essentially smaller
 $\chi^2\,$ than the commonly accepted one.

 \section{\bf 2. THE APT ESSENCE AND STRUCTURE}

\subsection{\bf 2a. \ Minkowskian And Euclidean Couplings
 \ $\alpha_M\,$ and $\alpha_E\,$}

 RG defined invariant coupling $\albar(Q^2)\,$ is a real
 function of space-like argument $Q^2\,.$ It effectively sums
 up UV logs into an expression with ghost. In the 1-loop
 QCD~case 
 \[\albars^{(1)}(Q^2)=\frac{\as}{1+\beta_0\,\as\,\ln(Q^2/\mu^2)}
 = \frac{1}{\beta_0\,\ln(Q^2/\Lambda^2)}\, \]
 Instead, in the APT scheme\cite{epj01}, we deal with differing
 ghost-free couplings
\begin{equation}\label{m1}
 \mbox{Minkowskian:}\quad \alpha_M^{(1)}(s)=\frac{1}
 {\pi\,\beta_0}\arccos\frac{L}{\sqrt{L^2+\pi^2}}\left.
 \right|_{L>0}=\frac{1}{\pi\beta_0}\arctan\frac{\pi}{L}\,;
 \quad \mbox{and} \eeq
 \begin{equation}\label{e1} \mbox{Euclidean:} \quad
 \alpha^{(1)}_E(Q^2)=\frac{1}{\beta_0}\left[\frac{1}{\ell}
 - \frac{\Lambda^2}{Q^2-\Lambda^2}\right]\,;\quad \ell=
 \ln\frac{Q^2}{\Lambda^2}\,,\quad
 L=\ln\frac{s}{\Lambda^2}\,.\eeq
\vspace{28mm}

\begin{center}
\begin{figure}[th] \unitlength=1mm
   \begin{picture}(0,20)                                   %
 \put(-55,60){\epsfig{file=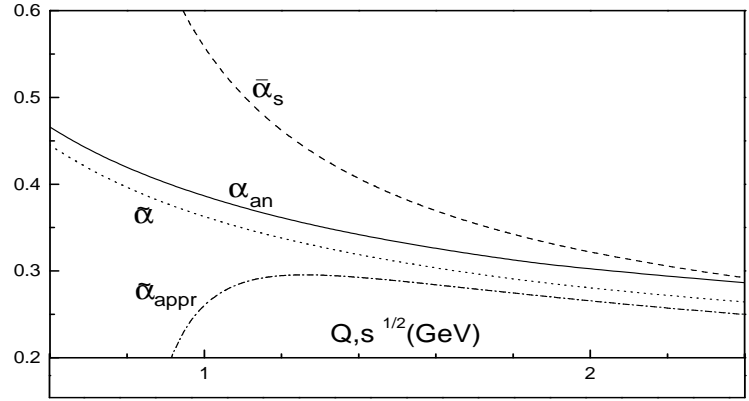,width=70mm,
 height=11.5cm,angle=-90}}
 \end{picture} \vspace{35mm}

 \parbox{15cm}{\caption{\sl\small{
 Comparison of usual QCD coupling \albars with Euclidean
 $\alpha_{an}=\alphaE\,$ and  Minkowskian one
 $\tildal=\alphaM\,$ in a few GeV region.} \label{fig1}}}
 \end{figure}
\end{center}

 On Fig.1 one can see\footnote{In this figure taken from our
 previous papers, a bit different notation
 $\alpha_{an}=\alphaE\,,\tildal=\alphaM$ is used. Here
 $\tildal_{appr}=\albars-\frac{\pi^2\beta_0^2}{3}\albars^3
 \,.$ All the curves are given in the 2-loop approximation for
 $\Lambda= 350\,\,\GeV\,.$} the comparison of \albars with
 $\alpha_E\,$ and $\alpha_M\,$ in the 1-2 \GeV \ region.

 Transition to the ``$s$ picture" performed first by contour
 integration by Radyushkin\cite{rad82}, Krasnikov and Pivovarov
 \cite{krapiv82}, (see also \cite{bjork89})\vspace{-2mm}
 \begin{equation}\label{R}
 \albars(Q^2) \to \frac{i}{2\,\pi}\int_{s-i\varepsilon}
 ^{s+i\varepsilon}\frac{d z}{z}\,\albars(-z) =\albarM(s)
 \equiv \mathbb{R}\:[\albars]\,(s)\,\end{equation}
 results in a ghost-free expression with $\pi^2$ terms summed.

  Reverse transformation $\mathbb{D}=[\mathbb{R}]^{-1}\,$
 \cite{rap96,prl97} yields\footnote{For its explicit form see
 below  eq.(\ref{D})} a ghost-free expression in the $Q^2\,$
 picture with subtracted singularity; see below eqs.(7) and (8).

 \subsection{\bf 2b. Minkowskian: \ $\pi^2\,$ Summation}
 Summation of $\,\pi^2\,$-terms by contour integration (3)
  for the 1-loop case results in
\begin{equation}\label{arccos}
\albars^{(1)}(Q^2)=\frac{1}{\beta_0 L}\to\albar_M^{(1)}(s)=
\frac{1}{\pi\,\beta_0}\arccos\frac{L}{\sqrt{L^2+\pi^2}}\,
 \equiv \, \agoth_1^{(1)}(s)\,, \eeq
\begin{equation}\label{arct}
 \left[\agoth_1^{(1)}(s)\right]_{L>0}=\frac{1}{\pi\beta_0}
\arctan\frac{\pi}{L}\,;\quad L=\ln\frac{s}{\Lambda^2}\,.\eeq
 This expression was first obtained by Radyushkin\cite{rad82}
 in the form (\ref{arct}). Later on, Jones and Solovtsov
 \cite{js95} considered the region $Q^2\leq\Lambda^2\,$ and
 proposed treating expression (\ref{arccos}) as a ghost-free
 Minkowskian effective coupling.\par
   At the same time, the procedure (\ref{R}) transforms
 square and cube of $\albars^{(1)}\,$ into ghost-free
 forms\cite{krapiv82}
 $$
 \agoth_2^{(1)}=\frac{1}{\beta_0^2\left[L^2+\pi^2\right]}\,,\ \
\agoth_3^{(1)}=\frac{L}{\beta_0^3\left[L^2+\pi^2\right]^2}\,,
 $$
 which are not powers of $\alpha_M^{(1)}(s)\,.$
 They are rather connected with (\ref{arccos}) by the
 iterative differential relation
\begin{equation}\label{iter1}
 \agoth_{k+1}(s)= -\frac{1}{k\,\beta_0}\,\frac{d\,\agoth_k(s)}
 {d \ln s}\,.\end{equation}

\subsection{\bf 2c.\ Euclidean:\ K\"allen-Lehmann Analyticity}
 APT uses imperative of the $Q^2\,$ analyticity\cite{bls59} in
 the form of the K\"allen--Lehmann spectral representation
 \footnote{In the form of the first of Eqs.(\ref{9}) For detail
 see Refs.\cite{rap96,prl97}.}. Being
 applied to the QCD one-loop case, it gives \vspace{-2mm}
\begin{equation}\label{7}
 \albars^{(1)}=\frac{1}{\beta_0\ell} \ \Rightarrow \
 \mathbb{A}\left[\albars^{(1)}\,\right]=\alpha^{(1)}_E(Q^2) =
\frac{1}{\beta_0}\left[\frac{1}{\ell}- \frac{\Lambda^2}
{Q^2-\Lambda^2}\right]=\acal_1^{(1)}(Q^2)\,.\eeq

 For coupling $\albars^{(1)}$ squared 

 \begin{equation}\label{8}
  \mathbb{A}\left[\frac{1}{\ell^2}\,\right] = \frac{1}
 {\ln^2(Q^2/\Lambda^2)}+\frac{Q^2\Lambda^2} {\left(Q^2-
 \Lambda^2\right)^2}=\beta_0^2\,\acal_2^{(1)}(Q^2)\,{
 \neq} \left(\beta_0\,\acal_1^{(1)}(Q^2)\right)^2\,.\eeq

 The Minkowskian and Euclidean ghost-free functions are
 related \cite{ms97}, \cite{epj01} by $\mathbb{D}\,$ and
 $\mathbb{R}\,$ transformations:
  $\acalk(Q^2)=\mathbb{D}\left[\agothk \right]\,,
 \quad \agothk(s)=\mathbb{R}\left[\acalk \right]\,.$
 Accordingly,  \vspace{-2mm} {
 $$\mathbb{D}\left[R(s)\,=\sum_k d_k\,\agothk(s)\right]
 \ \ \Rightarrow \ \ D(Q^2)=\sum_k d_k\,\acalk(Q^2)\,.$$}

\subsection{\bf 2d. Sketch Of The Global APT Algorithm}
  The most convenient form of the APT formalism uses a
 spectral density \ $\rho(\sigma)=Im\albars(-\sigma)\,$ taken
 from the perturbative input

\begin{equation}\label{9}
{\cal A}_k=\frac{1}{\pi}\int\limits_{0}^{\infty}
\frac{\rho_k(\sigma)\,d\sigma}{\sigma+Q^2}\,,\ \quad
\mathfrak{A}_k=\frac{1}{\pi}\int\limits^{\infty}_{s}
\frac{d \sigma}{\sigma}\,\rho_k(\sigma)\,. \eeq

 In the 1-loop case
  $$ \rho^{(1)}_1=\frac{1}{\beta_0\left[L_{\sigma}^2+\pi^2
 \right]}\,; \quad L_{\sigma}=\ln\frac{\sigma}{\Lambda^2}\,;
 \quad \rho_{k+1}^{(1)}(\sigma)= -\frac{1}{k\,\beta_0}\,
 \frac{d\,\rho_k^{(1)}(\sigma)} {d\,L_{\sigma}}$$
 These expressions were generalized for a higher-loop case and
 for real QCD with transitions across quark thresholds. This
 {\it global} APT was successively used for fitting of various
 data, e.g. for describing mass spectrum of light mesons
 \cite{ProspBald} and for description of pion formfactor
 \cite{Bak1}. Logic of the APT scheme is displayed\footnote{
 Here, the distance picture with functions $\alpha_D\,$ and
 $\left\{\aleph_k\right\}\,$ is mentioned. It is related with
 $Q^2$ picture by Fourier transfirmation $\mathbb{R}\,.$
 For detail, see Ref.\cite{dv04}.} in Fig.2.\vspace{23mm}

 \begin{center}
 \begin{figure}[th]  \unitlength=1mm
   \begin{picture}(50,20)                                   %
 \put(-17,-63) {\epsfig{file=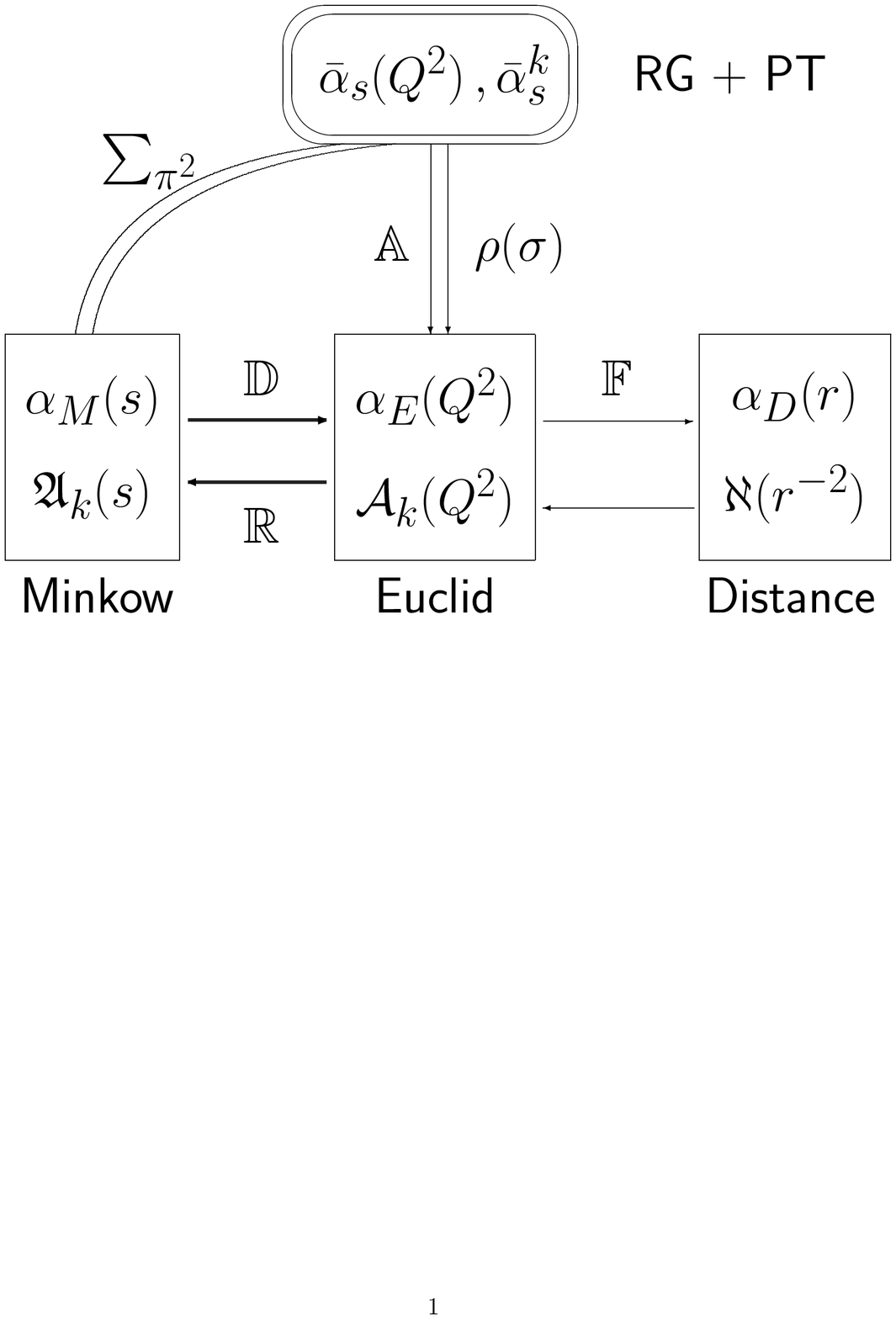, width=85mm}}
 \end{picture}

\parbox{15cm}{\caption{\sl\small{
 Logic of the APT scheme.} \label{fig2}}}
 \end{figure}\end{center}
 \vspace{-12mm}%

\section{\bf 3. THE APT RESUME}

\subsection{\bf 3a. Non-Power Ghost-Free Sets
 $\left\{{\cal A}_k\right\},\,\left\{{\mathfrak A}_k\right\}$}

 By construction, all APT expansion functions $\acal_k\,$ and
 $\agoth_k\,$ (for 2-loop etc. as well) are free of unphysical
 singularities and at weak-coupling limit tend to powers
 $\albars^k\,$ of common QCD coupling. On Fig.3 we demonstrate
  the behavior of the first three functions.\smallskip

Their more detailed properties can be described as follows:\smallskip

 {\bf I.}\ First ones, new couplings, $\alpha_E,\,\alpha_M\,:$\\
 $\diamondsuit$ are monotonic and IR finite\,,
  $\alpha_E(0)=\alpha_M(0 =1/\beta_0\simeq 1.4\,$\\
 $\diamondsuit$ in the UV limit
 $\,\sim 1/\ln x\/\sim\albars(x).$\\

{\bf II.} All the other functions \ $(k\geq 2)\,:$ \\
 $\heartsuit$ start from zero $\acal_k(0),\agoth_k(0)=0;$  \\
 $\heartsuit$ in the UV limit \ $\sim 1/(\ln x)^k\,\sim
    \albars^k(x)\,.$\\
 $\heartsuit$ 2nd ones, $\acal_2\,, \,\agoth_2\,$ obey max at
 $\sim \Lambda^2\,.$\\
 $\heartsuit$ Higher ones, $\acal_{k\geq 3}\,;\,\,
 \agoth_{k\geq 3}\,$ oscillate near $\Lambda^2$ with
 $\/k-1\/\,$ zeroes.  \vspace{-17mm}

\begin{center}
 \begin{figure}[th] \unitlength=1mm
   \begin{picture}(-3,68)                                   %
   \put(3,-8){\epsfig{file=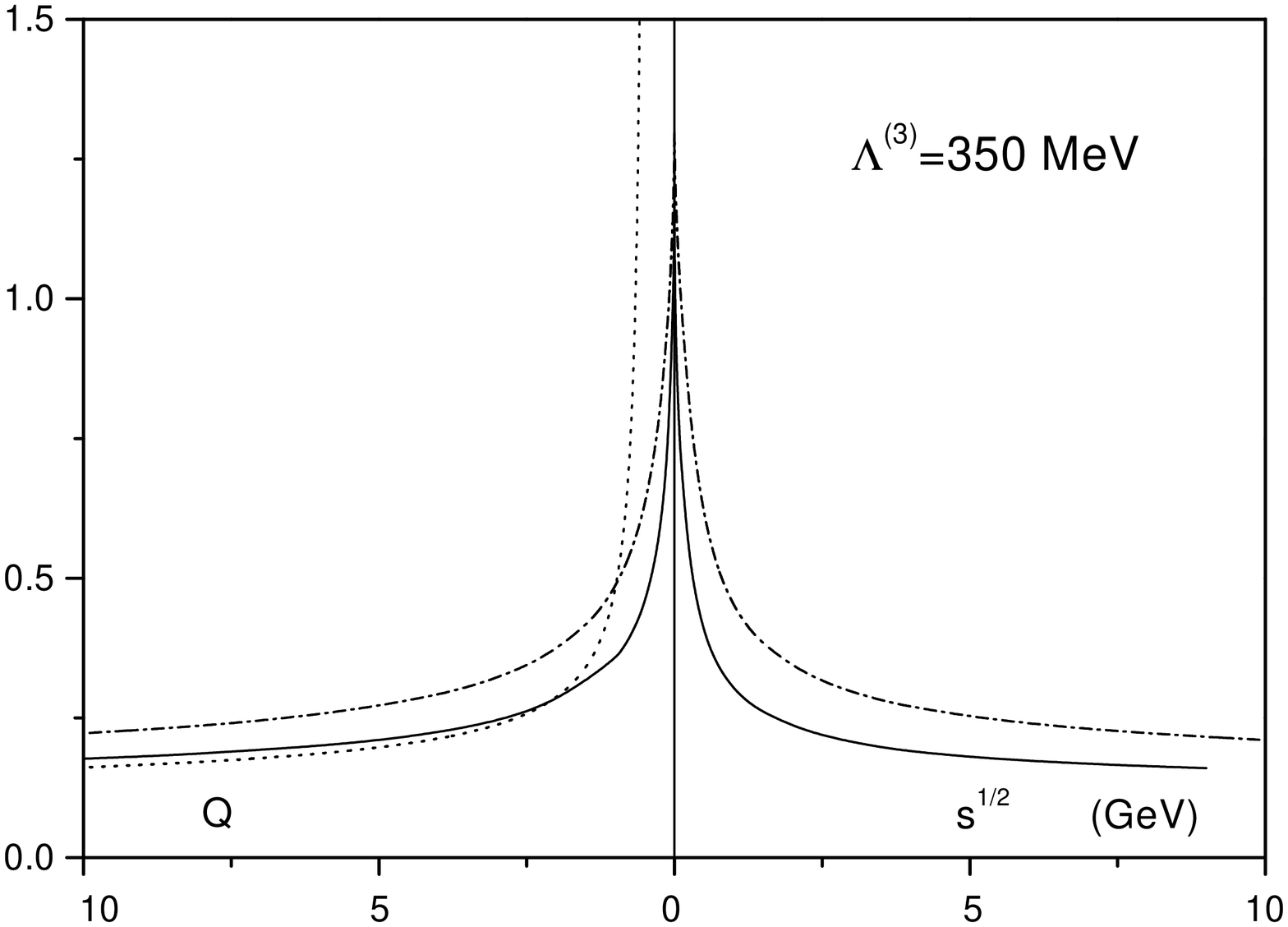,width=7.3cm,height=6.5cm}}
   \put(73,-8){\epsfig{file=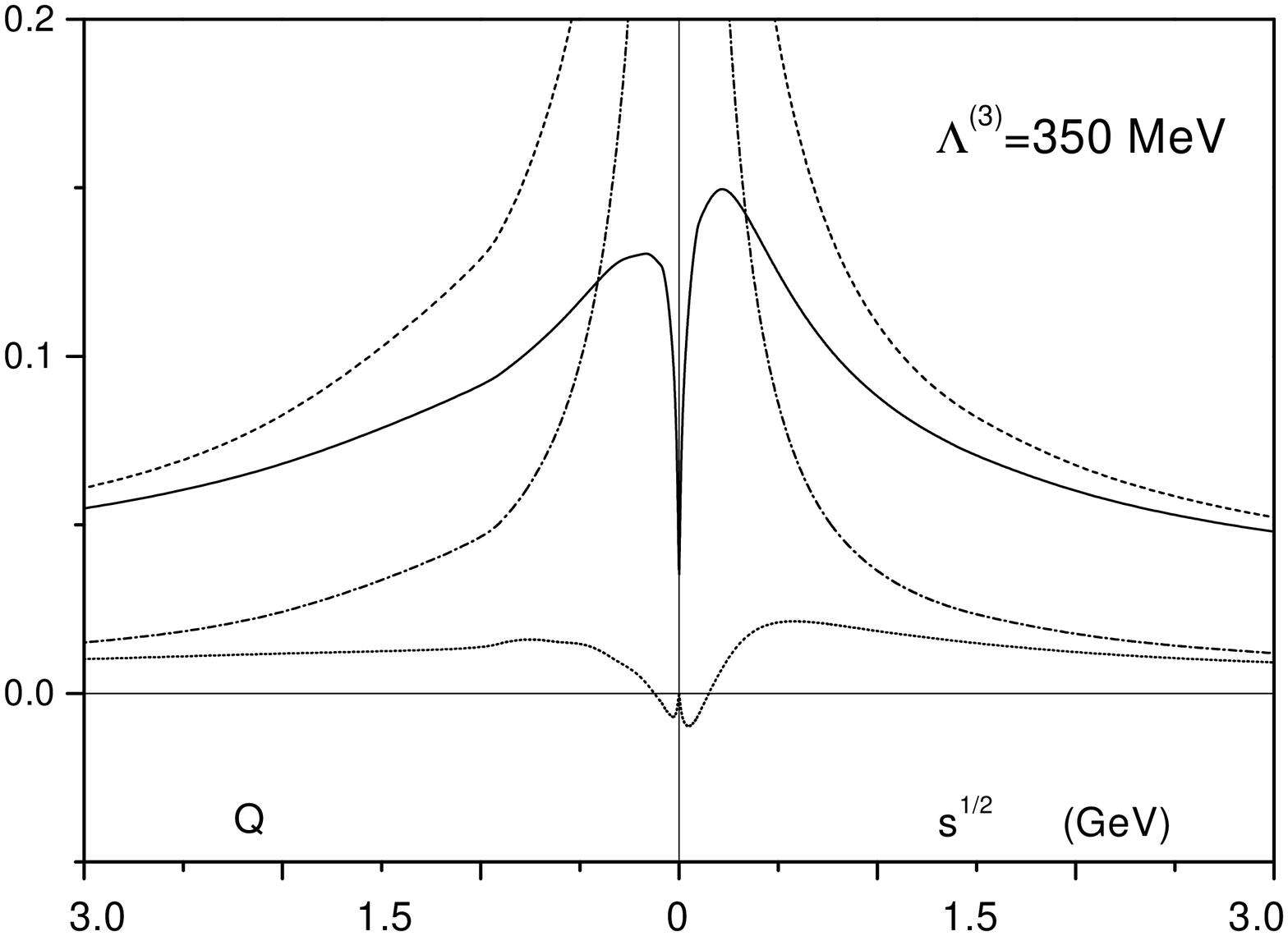,width=7.3cm,height=6.5cm}}       %
 \put(15,44){\bf a}                %
  \put(40.5,44){$\bullet$}
  \put(26,44){\small $\alpha_E(0)= \ \ \ \alpha_M(0)$}
   \put(25,31){\footnotesize $\bar\alpha_s^{(2)}(Q^2)$}
   \put(20.5,15){\small  $\alpha_E^{(1)}(Q^2)$}
   \put(29,5){ $\alpha_E^{(2,3)}$}
   \put(50,14){\small $\alpha_M^{(1)}(s)$}
   \put(42.5,4){ $\alpha_M^{(2,3)} $}
    \put(85,44){\bf b}
    \put(94,38){\footnotesize $\alpha_E^2(Q^2)$}
    \put(98,25){\small\bf $\acal_2$}
    \put(104,18){\footnotesize $\alpha_E^3$}
    \put(96,10){\small\bf ${\cal A}_3$}
    \put(122,33){\footnotesize $\alpha_M^2(s)$}
    \put(119,25){\small $\agoth_2$}
    \put(124,18){\footnotesize  $\alpha_M^3$}
    \put(117,10){\small ${\mathfrak A}_3$}
  \end{picture}  \vspace{16mm}

\centerline{ \parbox{12.2cm}{
\caption{\sl\footnotesize{\normalsize\bf a} -- Space-like
 and time-like APT couplings for 1-,2- and 3-loop case
 in a few $\GeV$ domain. \ \
 {\bf\normalsize b} -- ``Distorted mirror symmetry" for global
 expansion functions. All the solid curves here correspond to
 exact two--loop solutions $\,\acal_{2,3}\,$ and
 $\,\agoth_{2,3}\,.$ expressed in terms of the Lambert function.
 They are compared with powers of APT couplings $\alpha_E\,$ and
 $\alpha_M\,$ depicted by dotted lines.} \label{fig2} }}
 \end{figure} \vspace{-5mm}
\end{center}
 The last property results \cite{Bak1,ss98pl} in the reduced
 renormalization-scheme and higher loop sensitivity and better
 convergence \cite{Bak2} in the low-energy region, see below Sect.3b.

\subsection{\bf 3b. Non-Power Expansions: Quick Loop Convergence } 
 New effective couplings are related by integral transformations
 (\ref{R}) and
\begin{equation}\label{D}
 \alpha_E(Q^2) =Q^2\,\int^{\infty}_0 \frac{\alpha_M(s)\,d s}
 {(s+Q^2)^2}\equiv{\mathbb{D}} \left[\albar_M\right](Q^2)\,.\eeq
  The same transformations induce a nonpower structure
 \[{\mathfrak A_k}(s)\to{\cal A}_k(Q^2)\equiv{\mathbb{D}}
 \left[{\mathfrak A_k}\right](Q^2) \]
 of expansion functions for observables.

 Due to this, instead of the ``PDG--recommended" universal
 power-in-\albars expansion,
 \begin{equation}\label{pt} d_{pt}(Q^2/s)= d_1\albars(Q^2/s)+
 d_2\,\albars^2+{\bf d_3\:\albars^3 }\: \eeq
 one should use non-power expansions
\begin{equation}\label{d}
d_{\rm an}(Q^2)=d_1\,\alpha_E(Q^2) +d_2\,{\cal A}_2(Q^2)+
{\bf d_3\,{\cal A}_3(Q^2)}+ \dots\,,\end{equation}
\begin{equation}\label{r}
 r_{\pi}(s)= d_1\alpha_M(s)+d_2\,{\mathfrak A}_2(s)+{\bf
d_3\:{\mathfrak A}_3(s)}+ \dots\,.\end{equation}
  The numerical effect of this change is demonstrated in the
 Table 1. There, relative contributions in per cent for usual,
 PT 3-loop power-in-\alps expansions (\ref{pt}) are confronted
 with the APT ones (\ref{d}) and (\ref{r}). Besides, they are
 compared with the experimental error given in the last
 column in the same (i.e., in $\alps/\pi$) units.  

 \begin{center}
 {\bf Table 1.} {\small{Contributions in \%\%  of
 1-, 2-,3-loop  terms and data errors}} 
 \begin{tabular}{|l|l||c|c|c||r|r|r||r|}  \hline
\multicolumn{1}{|l|}{\slshape \phantom{aa}{\small Process}} &
\multicolumn{1}{l||}{\slshape {\small Energy}}
 &\multicolumn{3}{c||}{\slshape {\small PT (11)}}
 & \multicolumn{3}{r|}{\slshape{\small APT (12)/(13)}}
 & {\sf\footnotesize Exp Errors} \\    \hline\hline
{\footnotesize Bjorken SR}& 1.6\,GeV  & 55 & 26 & \bf{19} &
80 & 19 & \bf{1} \ & $\pm14$ \ \ \\ \hline
{\footnotesize GLS SRule }& 1.7\,GeV & 65 & 24 & \bf{11} &
 75 & 21 &\bf{4} \ & $\pm20$ \ \ \\ \hline\hline
{\footnotesize Incl $\tau$-decay}& 1.8 GeV & 55 & 29 &
 \bf{16} & 88 & 11 & \bf{1} \ & $\pm8$ \ \ \\ \hline
{\footnotesize$e^+e^-\to$hadr} & 10 GeV  & 96 & 8 & \bf{-4} &
92 & 7& \bf{.5}\ \  & $\pm27$ \ \ \\ \hline
$Z_o \to$ hadr. & 91  GeV & 99 & 3.7 & \bf{-2.3} & 97 & 3.5
& -\bf{.4} & $\pm4$ \ \  \\ \hline
\end{tabular}
\end{center}

 It follows that APT expansion converges much better than common
 PT one. Besides, the APT 3-loop term contribution is much less
 than data errors. Effective suppression of higher-loop terms
 yields also a reduced scheme \cite{Bak1} and loop dependence.

 All these nice features of APT are connected with due account
 for nonanalyticity with respect to usual expansion parameter,
 the coupling constant at $\alpha=0\,.$

 \subsection{\bf 3c. The QFT Nonanalyticity In Coupling}

  Here, we shortly remind a few general arguments on this
 non-analyticity.
 \begin{itemize}
 \item{\sf General Dyson \cite{dys52} argument in QED.} \ \
 Transition $\alpha \to -\alpha\,$ corresponds to $e\to i\,e\,;$
 it destroys Hermiticity of Lagrangian and the S-matrix unitarity.
 Hence, the origin $\alpha=0\,$ in the complex $\alpha$ plane
 can not be a regular point. \medskip

 \item {\sf RG\,+\,$Q^2$-analyticity arguments}.\ \ Combining
 the $\,Q^2\,$ analyticity for a photon propagator in QED with
 RG invariance, one could define \cite{dv76} the type of essential
 singularity at $\alpha=0\,$ as $\,\sim e^{-1/\alpha}\,.$ \medskip

 \item{\sf Functional integral reasoning.} \ By the method of
 functional-integral steepest descent for propagators, it was
 shown\cite{lip77} that expansion coefficients $c_n\,\alpha^n\,$
 at $n\gg 1$ behave like $\,c_n\,\sim n!\,n^m$ which
 corresponds\cite{bog77} to the same singularity
$\sim e^{-1/\alpha}$.
\end{itemize}

 \subsection{\bf 3d. \  Analytic approximations for 2-,
 3-loop \ \ $\agoth_k$ \ and \ $\acal_k\,$}
 Analytic expressions for 2-,3-loop APT Minkowskian $\agoth_k$ and
 Euclidean $\acalk\,$ couplings involving a special Lambert function
 \ $W_{-1}\,$ are rather cumbersome. Due to this, several analytic
 approximations for them were devised\cite{mss02,Bak2}. In addition,
 we can mention\cite{sz05} very simple "1-loop-like" model expressions
  with ``two-loop effective logs" $\ell_2\,,L_2\,$
 $$\agoth_1^*(s)=\frac{1}{\pi\,\beta_0} \arctan\frac{\pi}{L_2} \,,
 \quad \agoth_2^*=\frac{1}{\beta_0^2\left[L_2^2+\pi^2\right]} \,.\,
 \dots\, ;\quad L_2=L+b\,\ln L\,,\quad b=\frac{\beta_1}{\beta_0}\,,$$
 \[ \acal_1^{appr}=\frac{1}{\beta_0}\left(\frac{1}{l_2}-\frac{1}
 {\exp(l_2)-1}\right)\,,\,\dots \,\,;\quad
 \ell_2=\ell+ b\,\ln \ell\,, \]
 and modified parameter $\Lambda \to \Lambda_*=f(\Lambda)\,.$ Such
 analytic approximations, typically, could provide us with accuracy
 at the level of few \%\%  quite adequate to practical need. 

 \section{\bf 4. CONCLUSION}
\begin{enumerate}\setlength{\itemsep}{3.7pt}
 \item Numerous non-perturbative data (lattice simulations,
 Schwinger-Dyson eqs solution) reveal the ghost-free
 \albars behavior in low energy region with finite
  $\,\albars(0)\,$ value.
 \item The ''representation invariance" implies that functional
 expansions -- even in powers of some non-singular
 $\albar(Q^2/s)\,$ -- are not natural and should be changed for
 non-power perturbative-inspired expansions; this is essential
 in a few \GeV region.
 \item Hence, in this region:\\
 * the notion of a single universal effective charge \albars is
 not adequate, \\
 ** to correlate data, one needs two effective couplings
  \alphaEQ and \alphaMs.
 \item Instead of expansion (9.6) of PDG, one should use APT
 expansions eqs.(\ref{d}) and (\ref{r}) over sets of nonpower
 functions $\left\{A_k(Q^2)\right\}\,$ and $\left\{\agoth_k(s)
 \right\} \,.$.
\end{enumerate}

 \section{\bf ACKNOWLEDGEMENTS}

 The author is indebted to A.Bakulev, S.Mikhailov, A. Sidorov,
 N.Stefanis, O.Teryaev and A.Zayakin for useful discussions.
 This investigation was partially supported by grants RFBR
 05-01.00992 and Sc.Sch 2339.2003.2\,.

\end{document}